\documentclass{aastex}
\usepackage{spr-astr-addons}
\usepackage{url}

\RequirePackage{color}

\usepackage[hyperindex,breaklinks]{hyperref}
\begin{document}

\title{Search for exotic matter from gravitational\\microlensing observations of stars}
\shorttitle{Search for exotic matter}
\shortauthors{Bogdanov and Cherepashchuk}

\author{M.B.Bogdanov\altaffilmark{1}} \and
\affil{Chernyshevsky State University, ul. Astrakhanskaya 83, Saratov,
410012 Russia}
\author{A.M.Cherepashchuk\altaffilmark{2}}
\affil{Sternberg Astronomical Institute, Universitetskii pr. 13, Moscow,
119992 Russia}


\begin{abstract}

We consider small-scale spheroidal clusters of weakly
interacting massive particles in our Galaxy as non-compact
gravitational microlenses and predict the appearance of caustics in
the plane of a lensed source. The crossing of these caustics by a
lensed star can produce a large variety of light curves, including
some observed in actual microlensing events that have been
interpreted as manifestations of binary gravitational lenses. We
consider also observable effects during the gravitational
microlensing of stars of non-zero angular size with a given
brightness distribution across their disks by such an exotic
objects as natural wormholes and objects whose space-time
environment is described with the NUT metric. We demonstrate that,
under certain conditions, the microlensing light curves, chromatic
and polarizational effects due to the properties of the lens and
the star disk brightness distributions can differ considerably from
those observed for a Schwarzschild gravitational lens, so that
their analysis can facilitate the identification of such objects.
\end{abstract}


\section{Introduction}

The idea of detecting dark matter via its gravitational action on
the light arriving from distant stars suggested by \cite{bya69}
and \cite{pac86} was realized in observations of gravitational
microlensing of stars by massive compact halo objects in our Galaxy
\citep{alc93}. Various international teams - MACHO, OGLE,
PLANET, and others - have detected several hundred microlensing
events for stars in the Magellanic Clouds and in the bulge of the
Milky Way. However, the nature of the objects acting as
gravitational lenses remains unclear. According to the statistics
of observations by the MACHO team \citep{alc00}, the lens
masses are in the stellar-mass range. In two cases, the masses
exceed the limit for a neutron star, while the luminosities of
these lenses are no higher than that of the Sun \citep{ben02};
these objects are probably black holes.

A considerable fraction of the lenses may be dwarf stars in
our Galaxy. This is favored by statistical arguments \citep{kom95, ker97},
and also by the
analysis of light curves \citep{bog98, gau02}. In two cases
red dwarfs, that created the effect of
gravitational lens, were directly observed by the Hubble Space
Telescope \citep{alc01, koz07}. As noted
in the recent review of \cite{koc06},
the fraction of lenses from the dark matter can be sufficiently
small. However, uncertainties in the properties of the
gravitational lenses still admit a large variety objects of stellar
masses as candidates for this role.

Along with the black holes mentioned above, the possibilities
of the lenses being small-scale  clusters of weakly interacting
massive particles (WIMPs) \citep{gur97, saz96, zak99, bog01}
have also been considered.
Lenses of more exotic types of matter have also been suggested, in
particular, tunnels in space-time (wormholes) \citep{eir01, saf02, bog02},
as is predicted by general relativity. The papers \cite{rah03} and
\cite{rah04} considered the gravitational
microlensing of point sources by hypothetical objects whose local
regions of space-time are described with the NUT metric, and
compared theoretical light curves to observations.

The aim of the current paper is to study effects observed for
gravitational microlensing of stars by the exotic objects. It is
shown, in particular, that observations of stellar microlensing
evens can reveal fine photometric, chromatic, and polarizational
effects due to properties of the lens and lensed star: its angular
size, the brightness and polarization distributions across the
stellar disk, and the wavelength dependences of these
characteristics. Though the manifestations of such effects are
fairly small, their analysis can contribute to identification of
the lens type.

\section{Microlensing by Clusters of Weakly Interacting Massive Particles}

One approach to the solution of the dark-matter problem supposes
the existence of weakly interacting massive particles (WIMPs) that
formed at an early stage of the evolution of the Universe. Because
of their gravitational interaction, these particles should have
produced a hierarchical structure of clusters with certain scaling
properties. The masses of the smallest objects are of the order of
a solar mass. \cite{gur97} fitted the mass
density of the WIMPs in a small-scale  object with the following
function of the distance $r$ from the center of the cluster:

$$
\rho(r) =
\left\{
\begin{array}{cc}
\rho_0, & 0\leq r \leq r_c \\
\rho_0(r/r_c)^\alpha, & r_c\leq r \leq R_x \\
0, & r > R_x
\end{array}
\right.
\eqno(1)
$$

The exponent in (1) is also a scaling parameter, equal to $\alpha = 1.8$.
The radius of the core, which has a constant density $\rho_0$, is denoted
$r_c$ and can range from $0.05$ to $0.10$ of the cluster radius $R_x$.
According to \cite{gur97}, a typical cluster
has a mass of $M_x = 0.1 - 1.0\ M_{\odot}$ and a radius of $R_x
\sim 10^{14} - 10^{15}$ cm.

      The particles interact weakly both with each other and with
external matter, so that the detection of such clusters seems very
unlikely. However, \cite{gur95} pointed out that, if
such objects are located along the line of sight connecting an
observer and a normal star, they can deflect the light rays from
the star; in view of the masses of the clusters involved, this can
be considered gravitational microlensing. \cite{gur97} computed
flux-variation curves for the gravitational
microlensing of stars by non-compact lenses made up of the WIMP
clusters. The shapes of these curves can be rather similar to those
of a Schwarzschild lens, with a symmetric profile and a sharp
maximum, albeit with broader wings. In their comparison of these
curves with the first observational data obtained by the OGLE
group, \cite{saz96} showed that in two of
eight cases the observed light curves can be better described by a
non-compact lens model. Later, however, it was shown that the
broadening of the wings of these light curves can also be explained
in a simpler way - by either blending \citep{woz97}
or by the radiation of the lens itself, if it is a normal star
\citep{bog98}.

      \cite{zak99} analyzed a spherically symmetric nonsingular
model for a non-compact gravitational microlens, which describes
the expected density distribution in a WIMP cluster better than
earlier models. Under certain conditions, such a microlens can have
circular caustic curve in the plane of a lensed source and produce
a symmetric light curve with two maxima and a dip between them.
Similar curves can also be produced by binary gravitational lenses
and have already been observed in some microlensing events.

      In reality, the spherical symmetry of the density distribution
in a non-compact object can break down. One factor that can lead to
such a disruption of the symmetry is the tidal deformation of WIMP
clusters as a result of their gravitational interactions with each
other or with stars of the Galaxy. The possibility of such
deformations was pointed out by \cite{gur97}.
However, the probability of a close encounter involving appreciable
tidal forces is very low. Another factor - the oblateness of the
particle cluster by rotation - seems more likely.

      Rotation is a general property of systems that form as a
result of the gravitational contraction of matter. Rotation
explains, for example, the observed oblate shapes of elliptical
galaxies and globular star clusters. In the origin of rotation in
such systems, an important role should be played by gas-dynamical
effects, which are obviously absent in the case of weakly
interacting particles. There is, however, a mechanism capable of
inducing rotation of the particles exclusively via gravitational
interactions. This mechanism, first suggested by \cite{hoy49} and
then developed in more detail by \cite{pee69}, is based on the
mutual attraction of tidally deformed clusters. In this case, the
orbital momentum associated with the relative motion of the
clusters is transformed into rotational momentum of each of the
interacting clusters. In the early Universe, when WIMP clusters
formed, their spatial density was substantially higher than now.
Close encounters occurred rather frequently, and the Hoyle - Peebles
mechanism could be fairly efficient.

    The observed manifestations of a transparent gravitational
lens depend on the distribution of the surface mass density $\Sigma({\bf\xi})$,
where ${\bf\xi}$ is a vector defining the position of a point in the plane
of the lens. Here and below, we write vector names in bold font.
For the spherically symmetric law (1) in a coordinate system whose
origin coincides with the projection of the center of the particle
cluster onto the plane of the lens, $\Sigma({\bf\xi})$ is given by the integral
$$
\Sigma({\bf\xi})=2\int\limits_0^{\sqrt{R_x^2-\bf{\xi}^2}}
\rho(\sqrt{\bf{\xi}^2+z^2})\ dz.\eqno(2)
$$
It is obvious that the distribution of the surface mass density in
this case is rotationally symmetric. Further, we consider the
dimensionless density
$$
\kappa({\bf\xi})=\Sigma({\bf\xi})/\Sigma_{cr}, \eqno(3)
$$
where the so-called critical density $\Sigma_{cr}$ is determined by the
relation
$$
\Sigma_{cr}=\frac{c^2D_s}{4\pi GD_dD_{ds}}.\eqno(4)
$$
In (4), $c$ is the speed of light, $G$ the gravitational constant, $D_s$
the distance between the observer and lensed source, $D_d$ the
distance between the observer and lens, and $D_{ds}$ the distance
between the lens and source.

      Let us adopt the radius of the cluster $R_x$ as a scaling
parameter $\xi_0$ and introduce dimensionless vectors in the plane of
the lens ${\bf x} = {\bf \xi}/\xi_0$ and the plane of the source
${\bf y} = {\bf \eta}/\eta_0$, where $\eta_0=\xi_0 D_s/D_d$, $\xi_0 = R_x$,
${\bf x}(x_1, x_2)$, and ${\bf y}(y_1, y_2)$. We use ${\bf \alpha }({\bf x})$
to denote the normalized vector ${\bf \alpha}^0(\xi_0{\bf x})$:
$$
{\bf \alpha }({\bf x})= \frac{D_dD_{ds}}{\xi_0 D_s} {\bf \alpha}^0(\xi_0{\bf x}).
$$
The gravitational lens equation \citep{koc06, zak97} can then be written
in the form
$$
{\bf y} = {\bf x} - {\bf \alpha }({\bf x}). \eqno(5)
$$

      The possibility of WIMP clusters being oblate by rotation can
hardly provoke any fundamental objections. However, the questions
of the oblate cluster's shape and density distribution have no
simple answers. The branch of celestial mechanics that studies the
equilibrium shapes of rotating bodies assumes them to be comprised
of a viscous incompressible fluid, which is clearly inappropriate
for systems of weakly interacting particles. Nonetheless, some
conclusions may be of interest for our case as well. Given the
presence of a small dense core and a rapid decrease of the particle
density (1) with distance from the cluster center, we can describe
this density distribution to a first approximation using the Roche
model, which assumes that the entire mass of the rotating body is
concentrated at its center \citep{sub49}. This type of model has
been used earlier to study planetary atmospheres and proto-
planetary disks. The oblateness of the marginally stable surface of
the Roche model was fairly small, with a polar-to-equatorial-axis
ratio equal to $2/3$ \citep{sub49}.

   Models with a nonsingular power-law elliptical distribution of
the surface mass density have been widely used in studies of
transparent gravitational lenses:
$$
\kappa({\bf x})= \frac{q}{(u^2+x_1^2+x_2^2/k^2)^n}. \eqno(6)
$$
This model has a constant-density core of radius $u$ at the center.
The contours of equal surface density are ellipses with fixed minor
to major semiaxis ratio $k = b/a$. Models of the form (6) are
believed to adequately describe the surface mass density
distributions in galaxies for a wide range of degrees of
oblateness.

      We first analyzed the case of a non-rotating cluster and tried
to fit the density distribution of a WIMP cluster (3) with $r_c =
0.07 R_x$ and normalization $\kappa({\bf 0}) = 1$ using the model (6) with $k = 1$,
choosing model parameters to minimize the sum of squared residuals.
The solution of the corresponding optimization problem yielded the
best-fit parameter values $q = 1.04$, $u = 0.09$, and $n = 0.68$. The
surface density distribution for model (6) with these parameters is
a good fit to the initial distribution. We then assumed that, as
the particle cluster rotates, the parameters $q$, $u$, and $n$ remain
constant and the ellipses of equal density experience oblateness
equal to that of the marginally stable surface in the Roche model:
$k = 2/3$.

     It is well known that a transparent gravitational lens with
surface density distribution (6) can have caustics in the source
plane, which, when crossed by a lensed star, can produce a large
variety of light curve shapes. For this to happen, it is sufficient
for a point ${\bf x}$ with $\kappa({\bf x}) > 1$ to exist in the plane of the lens
\citep{koc06, zak97}. This
condition can be satisfied by the WIMP clusters considered. Let us
fix $R_x = 10^{14}$ cm and $M_x = 0.5\ M_{\odot} = 10^{33}$ g. It then follows
from (1)
that $\rho_0 = 1.8 \times 10^{-8} g/cm^3$, and the computation of integral (2)
yields a surface density of $\Sigma({\bf 0}) = 3.4 \times 10^{5} g/cm^2$ at the lens
center. We find from (4) that the critical density is $\Sigma_{cr} = 5.3 \times
 10^{4} g/cm^2$ and $\Sigma_{cr} = 1.7 \times 10^{5} g/cm^2$ for typical
cases of lensing of
stars in the Large Magellanic Cloud by an object located in the
Galactic halo ($D_d = 8$ kpc, $D_s = 46$ kpc, and $D_{ds} = 38$ kpc) and
lensing of Galactic bulge stars ($D_d = 4$ kpc, $D_s = 8$ kpc, and $D_{ds} = 4$
kpc), respectively. Thus, the required excess of the density over
its critical level is quite possible.

      Let us assume for the sake of simplicity that the rotational
axis of the cluster lies in the plane of the lens and that the
major axes of the ellipses of equal surface density form intervals
on the $x_1$ axis. It is necessary however to note that this alignment of cluster
maximizes the asymmetry of the lens and the size of the caustics.
\cite{bar98} has developed an efficient algorithm
for solving the gravitational-lens equation (5) using the density
distribution (6) and written a FORTRAN code implementing this
algorithm. For a given vector ${\bf x}$, the code computes the components
of the deflection angle vector ${\bf\alpha}({\bf x})$ and the elements of the Jacobi
matrix $A_{i,j} = \partial y_i/\partial x_j$ . Our computations for model (6) with the
parameters adopted here demonstrated the existence of critical
curves in the plane of the lens, defined by the condition $det A =
0$. The corresponding caustics in the plane of the source are shown
by the solid curves in Figure 1. When a point source crosses the
caustic, the gravitational lens amplification factor $\mu = 1/|det A|$
can formally approach infinity. For a source with a finite angular
size, $\mu$ is finite but can achieve rather large values.

\begin{figure} [t]
\plotone{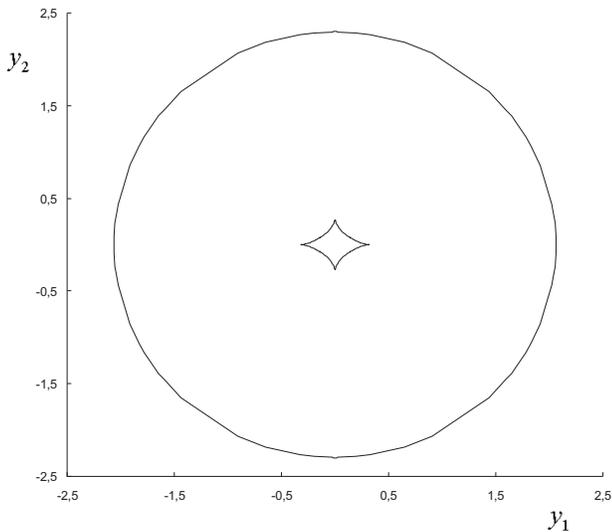}
\caption[]{
Caustic curves of the non-compact WIMP lens in the
plane of the source
}
\label{f1}
\end{figure}

      For the adopted size of the WIMP cluster, the unit scale
length in the plane of the lens for $D_d = 4$ kpc corresponds to an
angle of $0^{\prime\prime}.0017$, which is much smaller than the resolution of
current telescopes. As a result, the object should show up as a
microlens, with the total flux from all its images being observed.
Due to the orbital motion of the observer, particle cluster, and
lensed star, the observed flux should vary with time. Exact
computation of the light curves is computer intensive. However,
it is easy to obtain a qualitative
description of the observational picture. When a source of finite
angular size crosses the outer caustic, two additional source
images should appear and the observed flux should increase.
However, the flux amplification factor on this caustic is
comparatively small, and the caustic crossing could escape
detection. When the source crosses the inner caustic, two more
source images appear, resulting in a sharp increase of the flux
followed by a slower decrease. At the second crossing of the inner
caustic, the two images merge and disappear, inverting the pattern
of the flux variations. When the line of the relative motion of the
lensed star in Fig. 1 passes far from the cusp and is parallel to
the coordinate axes when it crosses the inner caustic, the light
curves have symmetric profiles with two maxima and a dip at the
center. Such curves can be realized in the noncompact microlens
model analyzed by \cite{zak99}. If the star moves along a line
at some angle to the $y_1$ axis, either the first or second crossing
of the inner caustic should take place closer to the corresponding
cusp. As a result of the difference in the amplification factors,
the light curve becomes asymmetric and either its first or second
maximum becomes stronger. Similar light-curve shapes can appear in
the cases of binary gravitational lenses and have been observed in
actual microlensing events. We must therefore bear
in mind the possibility that at least some such events could in
reality be associated with WIMP clusters. When the star goes beyond the outer
caustic, comparative modest flux variations will again be observed
as a flux increase followed by a more abrupt decrease. If the
lensed star does not cross the outer caustic, the flux-variation
curve is symmetrical and has a single maximum.

      The microlensing pattern considered above depends on a large
number of parameters. Obviously, it is possible to obtain a large
variety of light curves by changing the lens orientation and
varying the parameters for the density distribution and the
relative motion of the lensed star. Represents the interest to examine
quantitative a question how well a light curve for binary
lens can be fitted by curve for the model of noncompact lens.

\section{Spatial - Temporal Tunnels as Gravitational Microlenses}

As was shown by \cite{mor88}, \cite{hoc97a}, and
\cite{hoc97b}, the equations of
general relativity allow solutions in the form of two regions in
space-time connected by a so-called spatial-temporal tunnel or
natural wormhole. Such a tunnel can exist only if it is filled with
a certain type of exotic matter with a negative energy density and,
therefore, a negative mass. It was first suggested by \cite{kim96}
that such an object could be manifest as a gravitational
lens capable of forming multiple images of a distant source of
radiation. Under certain conditions, the angular distance between
the images can become smaller than the resolving power of
telescopes, and the observer will detect only the combined flux
from all the images of the source. In this way, the wormhole will
be manifest as a gravitational microlens with negative mass. Such a
microlens modeled as a negative point mass was considered by \cite{eir01},
and \cite{saf02}
using the commonly applied approximations of geometric optics and
small deflection angles. Like an ordinary gravitational lens, it is
basically specified by the angular radius of the Einstein cone $p_0$,
which can be expressed as
$$
p_0^2=\frac{4G|M|}{c^2}\frac{D_{ds}}{(D_{ds}+D_d)D_d}, \eqno(7)
$$
where $M$ the mass of the lens. Unfortunately, any reliable estimations
of mass of wormholes and hence $p_0$ until today are absent.

The lens equation indicates that its
properties differ dramatically from those of an ordinary positive-
mass Schwarzschild lens due to the presence of a circular caustic
with angular radius $2p_0$ in the source plane. When a point source is
located an angular distance $p < 2p_0$ from the lens, it is no longer
observed, and the detected flux decreases to zero. When $p > 2p_0$,
two images of the source arise, located on one side of the lens
between the directions towards the lens and the source. One of the
images is always inside the Einstein cone, while the second is
outside it. As the source approaches the caustic from the outer
side, the images merge and disappear when $p = 2p_0$. The
amplification factor for the flux from a source with an infinitely
small angular size in the region $p > 2p_0$ is
$$
A(u)=\frac{u^2-2}{u\sqrt{u^2-4}}, \eqno(8)
$$
where $u = p/p_0$; at a caustic crossing, $A(u)$ formally tends to
infinity. Relation (8) differs from the corresponding formula for a
Schwarzschild lens \citep{koc06, zak97} only in the signs of the
numerical terms in the numerator and denominator.

      In the case of lensing of a source with a finite angular size,
the flux will increase as it approaches the caustic, then sharply
decrease and completely disappear. In the course of the egress of
the source from the inner region of the circular caustic, the
pattern will change to the opposite. For a source with a circularly
symmetrical brightness distribution, the light curve will always be
symmetrical relative to the time of its closest approach to the
lens. If the source does not cross the caustic during its relative
motion, the light curve displays one maximum and its shape is close
to those of observed stellar microlensing light curves. As it was
shown by \cite{bog02}, if a point source moves
outside the circular caustic, photometric observations cannot
distinguish light curves due to a wormhole and a positive-mass
gravitational lens. However, it is apparent that the amplification
factors of the lenses can be equal only for one value of the impact
parameter, so that the light curves for extended sources will
differ.

      Images and light curves of extended sources formed by a
negative-mass lens were calculated by \cite{saf02} using the method
of ray tracing and the solution of the lens
equation on a $5000 \times 5000$ count grid. If we consider only
photometric effects, the problem is simplified by the fact that the
amplification factor for the flux from a source area element will
be specified by (8). The problem is thus similar to the calculation
of light curves for the case when a source is lensed by a linear
caustic \citep{bog00}. In this case,
photometric effects depend only on the one-dimensional projection
of the source brightness distribution onto the axis perpendicular
to the caustic usually called the strip brightness distribution. In
our case, the lens possesses a circular caustic, and the light
curve $I(p)$ as a function of the angular distance from the lens $p$
will depend on the one-dimensional brightness distribution $B_p(r)$,
which is the source brightness integrated over an infinitely narrow
circular strip with radius $r$ concentric to the caustic. Within this
strip, the lens amplification factor (8) remains constant. If the
angular size of the source is much smaller than $p_0$, the curvature
of the caustic can be neglected, and the distribution $B_p(r)$
coincides with the strip brightness distribution. As in the
previous case of a linear caustic, the influence of the singularity
of the integral in the flux calculation can be removed by
corresponding selection of the grids for the variables $p_i$ and $r_i$
\citep{bog02}.

      Further, we will assume the lensed source to be a star and the
brightness distribution over the stellar disk to display circular
symmetry. In this case, the detected flux will depend only on the
angular distance between the center of the stellar disk and the
lens. For most stars, the limb darkening is described by the linear
law:
$$
b_{\lambda}(\mu) = b_{\lambda}(1) ( 1 - x_{\lambda} + x_{\lambda} \mu ),
\eqno(9)
$$
where $\mu$ is the cosine of the angle between the line of sight and
the normal to the stellar surface and $x_{\lambda}$ is the limb-darkening
coefficient, which depends on the wavelength $\lambda$. The value $\mu$
corresponding to the angular distance from the center of the star $\rho$
is related to it by the expression:
$$
\mu=\sqrt{R^2-\rho^2}/\rho ,
$$
where $R$ is the angular radius of the star. Further, we will assume
that the brightness at the center of the star can be written
$$
b_{\lambda}(1) = 1/\pi R^2(1- x_{\lambda}/3) ,
$$
which normalizes its flux to unity.

    Let us introduce a polar coordinate system $(r, \varphi)$ with its
origin coincident with the lens. If the center of the stellar disk
is located an angular distance $p$ from the lens, the integrated
brightness of the circular strip of the disk specified by the
radius $r$ is
$$
B_p(r)=2\int\limits_0^{\varphi_m}
b_{\lambda}(\rho)r\ d\varphi, \eqno(10)
$$
where $ \rho^2 = r^2 + p^2 - 2rp \cos \varphi $  and an upper limit for the integral
can be derived from the condition $R^2 = r^2 + p^2 - 2rp \cos \varphi_m $ . Taking
into account the caustic amplification factor (8), the detected
radiation flux is
$$
I(p)=\int\limits_{p-R}^{p+R}
A(r/p_0)B_p(r)\ dr   , \eqno(11)
$$
where $A(r/p_0) = 0$ for $r < 2p_0$ .

For the purpose of the estimation of the effect of
finite angular size of the source we chose a $K0V$ star, and assumed
that its limb-darkening coefficient is $x_V = 0.702$ in the $V$ band
\citep{rub90}. The light curves were calculated using (9),
(10) and (11) and canonical-dissection grids with a step
sufficiently small that the relative error would not exceed $10^{-4}$.
Since the flux depends only on the angular distance $p$, the light
curve can be calculated for a single value of the impact parameter
$p_m = 0$; we can find the counts for other values via interpolation.
We used a cubic spline interpolation, which provided the necessary
accuracy. We chose the value $R = 0.10$ for the comparison with the
calculations of \cite{eir01} (further, we will
measure all angles in units of $p_0$).

      The light curves were calculated for the impact parameters $p_m
= 0, 1.0, 1.9$, and $2.1$. Further, we will measure time in units of
the time required to cross the angular radius of the Einstein cone
$t_E$ . Figure 2 presents these results. Since all the curves are
symmetrical about the time of closest approach to the lens, the
figure shows only their positive branches. Our calculations are
consistent with the results of \cite{eir01},
confirming the validity of the technique used. We can see from Fig.
2 that, when the star does not cross the circular caustic, the
light curve displays a single maximum and its shape is similar to
that for a Schwarzschild lens.

\begin{figure} [t]
\plotone{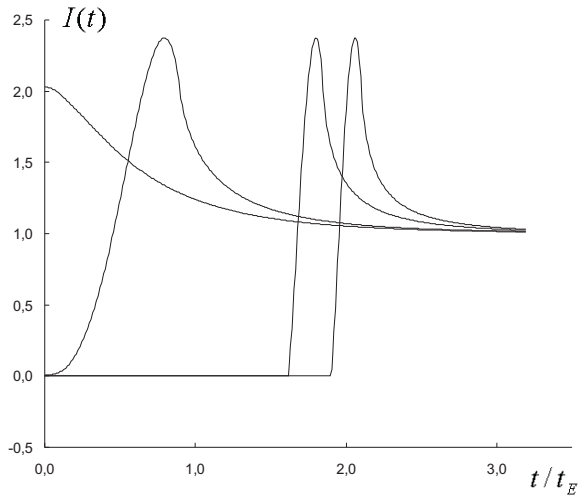}
\caption[]{
V light curves for the microlensing of a star with
angular radius 0.10 by a wormhole; time is measured in units
of the Einstein cone angular radius crossing time. The
positive branches of the curves for the impact parameters $p_m =
2.1, 1.9, 1.0,$ and $0$ are indicated
}
\label{f2}
\end{figure}

      When a star crosses the circular caustic, characteristic
symmetrical light curves with a central depression are formed. They
differ markedly from observed stellar microlensing light curves,
and could potentially provide unambiguous evidence for the presence
of a negative-mass gravitational lens. However, in the study by
\cite{boz02} of microlensing of stars by a compact object
surrounded by an extended gaseous envelope, due to the extinction
of the stellar radiation by refraction and Rayleigh scattering in
the envelope, the light curve can acquire a shape close to that
characteristic of a negative-mass lens. One possibility for
discriminating between these two types of lens is provided by
analyses of chromatic effects associated with the lensing.

      The basic property of gravitational lenses of both positive
and negative masses is that their parameters are independent of
wavelength. However, the amplification factors are different for
radiation from different regions of the stellar disk. Due to the
wavelength dependence of the brightness distribution, this results
in chromatic effects during the microlensing. The possibility of
observing these effects for ordinary gravitational lenses has been
considered, for example, in papers of \cite{bog95}, \cite{loe95}, and
\cite{han00}. Such an analysis for a negative-mass lens was
carried out in paper of \cite{eir01} by
calculating the $U$ and $I$ light curves for a $K$ giant with $T_{eff} = 4750\
K$ and $R = 0.10$, assuming that the limb-darkening coefficients were
$x_U = 1.050$ and $x_I = 0.503$. The limb darkening coefficients $x_{\lambda}$
were determined from calculations of stellar atmosphere models. In this
case, $x_{\lambda}$ can be estimated in different ways, based on either
conservation of the total flux or the best consistency between (9)
and the calculated data \citep{rub90}. The first method can
formally give a value $x_{\lambda}> 1$ . However, it is physically
meaningless to use this value to calculate the brightness
distribution (9), as was done in paper of \cite{eir01}, since this
yields negative brightness at the edge of the stellar disk.

     We calculated the variations of $V - I$ for microlensing of a $K0V$
star by a negative-mass object, since exactly these photometric
bands are most frequently used in searches for gravitational-
microlensing events. The limb-darkening coefficients $x_V = 0.702$ and
$x_I = 0.433$ were taken from paper of \cite{rub90}. As for the
calculation of the photometric effects, we adopted $R = 0.10$ and $p_m
= 0, 1.0, 1.9, 2.1$. Figure 3 presents the resulting positive
branches of the color-variation curves. As expected, the maximum
variation $\Delta (V - I)$, which corresponds to reddening, occurs when the
stellar disk is almost entirely inside the circular caustic.
However, the radiation flux is very small in this case (see Fig. 2)
and cannot be detected. We restricted the calculations to the
region in which the $V$ flux exceeds $0.001$ of its initial value. In
the course of the passage of the stellar disk across the caustic,
$\Delta (V - I)$ changes sign twice: after reddening, the star becomes bluer,
then redder again. However, in the region where the flux is
appreciably different from zero, the chromatic effects are small
and can be detected only with rather high photometric accuracy. If
the star does not cross the circular caustic, the color varies in
the same way as for a Schwarzschild lens \citep{bog95} and reddening is
always observed.

\begin{figure} [t]
\plotone{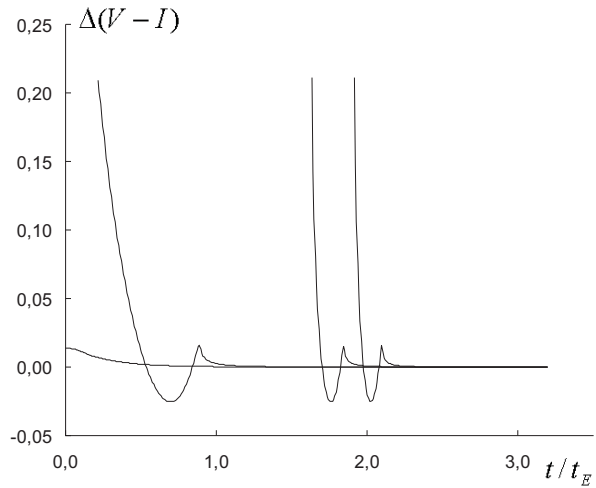}
\caption[]{
Variation of the color index $\Delta(V-I)$  for the
microlensing of a star by the wormhole shown in Fig. 2. The
units along the horizontal axis are the same as in Fig. 2
}
\label{f3}
\end{figure}

      Thus, in the same way as for ordinary gravitational lenses,
chromatic effects during the microlensing of a star by a wormhole
are second-order compared to photometric effects. However, their
detection could make it easy to discriminate between this exotic
object and a compact body with an extended gaseous envelope. In the
latter case, the chromatic effects will be very different in both
their values and character, due to the strong wavelength dependence
of the refraction and, especially, of Rayleigh scattering ($\propto \lambda^{-4}$).

      Analogous to chromatic effects, gravitational microlensing of
stars gives rise to variable polarization of the observed
radiation, due to the character of the polarization distribution
over the stellar disk. The stellar radiation becomes polarized as a
result of scattering in the stellar atmosphere. \cite{cha53} and
\cite{sob56} demonstrated that the radiation emerging
from a plane-parallel Rayleigh atmosphere is partially polarized,
with the direction of the electric vector being perpendicular to
the plane containing the line of sight and the normal to the
stellar surface. In ordinary observations of stars, the detected
total radiation flux from the entire stellar disk is unpolarized
due to the disk's symmetry. The origination of partial polarization
of the total flux is associated with a violation of the disk's
symmetry due to rapid rotation of the star, tidal deformation, the
presence of hot spots and/or an asymmetrical envelope, or a
combination of these factors. It is apparent that the action of a
gravitational lens also violates the disk's symmetry. The type and
degree of polarization observed during microlensing of a star by a
Schwarzschild lens were considered in papers of \cite{sim95} and
\cite{bog96}.

      To analyze the linear polarization that originates during the
propagation of radiation in the atmosphere of a star, it is
convenient to introduce the intensities $b_l$ in the plane, which
contains normal to the stellar surface and line of sight, and $b_r$ in
the direction perpendicular to this plane. For simplicity, we will
assume that the lensed star has a plane-parallel Rayleigh
atmosphere. In this case, the exact brightness distributions $b_l$ and
$b_r$ over the stellar disk are presented by \cite{cha53}. For
our estimates, it is more convenient to use the approximate
formulas derived by \cite{boc83}:
$$
b_r+b_l=\frac{1+16.035\ \mu + 25.503\ \mu^2}{1+12.561\ \mu +0.331\ \mu^2} ,
\eqno(12)
$$
$$
\frac{b_r-b_l}{1-\mu}= \frac{0.1171+3.3207\ \mu +6.1522\ \mu^2}
{1+31.4160\ \mu + 74.0112\ \mu^2} , \eqno(13)
$$
where, as in (9), $\mu$ is the cosine of the angle between the line of
sight and the normal to the stellar surface. Relations (12) and
(13) can be used to estimate the intensities of the polarized
components, so that the relative error of the degree of
polarization $P = (b_r - b_l)/(b_r + b_l)$ compared to the exact solution
for any $\mu$ does not exceed $0.001$.

      Let us bring the origin of the polar coordinate system $(\rho,\theta)$
to the center of the stellar disk, so that the main axis from which
we measure the angle $\theta$ will be directed towards the gravitational
lens, located an angular distance $p$ from the coordinate origin. Due
to the circular symmetry of the initial brightness distributions $b_l$
and $b_r$, it is apparent that the flux and degree of polarization of
the detected radiation will depend only on $p$. Let $b_L(\rho,\theta)$ and
$b_R(\rho,\theta)$ be the intensities at the point of the visible stellar
disk with the coordinates $(\rho,\theta)$ for observations through
polarizers oriented parallel and perpendicular to the direction
towards the lens. Then, according to the transformation law for the
Stokes parameters (and, accordingly, for the intensities $b_l$ and
$b_r$), we have:
$$
b_L(\rho,\theta)= b_r(\rho)\sin^2 \theta + b_l(\rho) \cos^2 \theta
, \eqno(14)
$$
$$
b_R(\rho,\theta)= b_l(\rho)\sin^2 \theta + b_r(\rho) \cos^2 \theta .
\eqno(15)
$$
The coordinate system $(\rho, \theta)$ is related to the system $(r, \varphi)$
introduced above: $r \sin \varphi = \rho \sin \theta$. Taking this into account
and substituting the brightness distributions (14) and (15) into (10)
and (11), we can calculate the flux from the star with the
polarizer oriented in the parallel ($I_L(p)$) and perpendicular
($I_R(p)$) directions relative to the lens. The degree of polarization
$P(p)$ when the center of the star is an angular distance $p$ from the
lens will be
$$
P(p)= \frac{I_R(p) - I_L(p)}{I_R(p) + I_L(p)} . \eqno(16)
$$

Figure 4 presents the positive branches of the polarization
curves for the microlensing of a star with a Rayleigh atmosphere
calculated for the same lensing parameters as in the previous case.
As previously, we will consider only those sections of the curves
for which the flux is no lower than $0.001$ of the initial value.
Since, like the chromatic effects, the polarization effects of
microlensing are due to differences in the amplification of the
radiation from the central and limb regions of the stellar disk, it
is not surprising that the shape of the polarization curves
essentially coincides with that of the color-index variation curves
(see Fig. 3). During the passage of the caustic across the stellar
disk, the degree of polarization changes sign twice. According to
(16), this means that the plane of the electric vector is first
perpendicular ($P > 0$), then parallel ($P < 0$), and finally
perpendicular ($P > 0$) to the direction towards the lens.

\begin{figure} [t]
\plotone{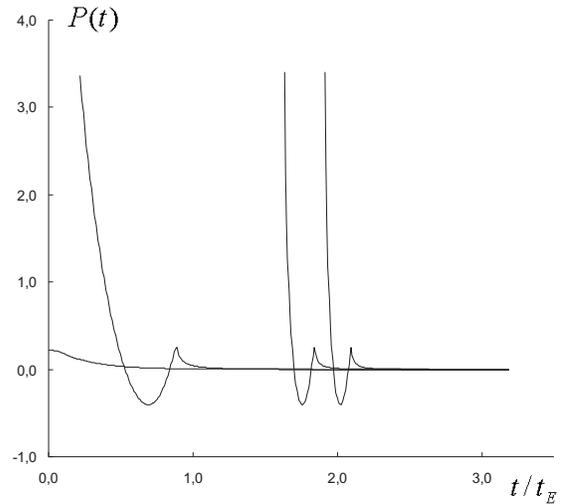}
\caption[]{
The variation of the degree of polarization $P(t)$ (in percent) for
curves of microlensing of a star by the wormhole shown in Fig.
2. The units along the horizontal axis are the same as in Fig.2
}
\label{f4}
\end{figure}

      When the star does not cross the circular caustic, the degree
of polarization is always positive, and the plane of the electric
vector is always perpendicular to the direction from the center of
the star towards the gravitational lens. Thus, with sufficient
observational accuracy, it may be possible to detect the rotation
of the plane of polarization as the star moves relative to the
lens. At the time when the maximum degree of polarization is
attained, which coincides with the maximum of the flux from the
lensed star, the plane of polarization will be specified by the
position angle of the relative motion of the star and lens. The
direction of the motion remains unspecified, due to the symmetry. A
similar pattern is observed for a Schwarzschild lens \citep{bog96}.

      For a star with a pure Rayleigh atmosphere, the degree of
polarization is small, and its maximum, reached at the very edge of
the stellar disk, is $11.7\%$. Since the albedo of a singular
scattering under the conditions in stellar atmospheres is always
smaller than unity, we expect that the polarizations of real stars
will be lower, so that the calculated values correspond to upper
limits. In addition, the Nagirner effect may also be observed
\citep{nag62}, which is manifest by a rotation of the plane of
polarization with motion from the edge of the stellar disk. The
influence of the true absorption on the polarization of the
radiation emerging from the atmosphere of the star assuming a
plane-parallel geometry was considered in detail by \cite{boc85}.
In the Wien region of the
spectrum, $P$ can appreciably exceed the value for a Rayleigh
atmosphere. The degree of polarization for long wavelengths is
smaller, but remains appreciable. For late-type dwarfs, which
statistically are most likely to be lensed, an additional
contribution to the increase of $P$ may be introduced by Rayleigh
scattering on molecules. Overall, for real stars, the order of
magnitude of the calculated polarizations is expected to remain
valid.

      Thus, both chromatic and polarization effects during the
microlensing of a star by a wormhole are small. However, the
accuracy of current polarimetric observations considerably exceeds
photometric accuracy; for sufficiently bright objects, $P$ can be
determined with an error not exceeding 0.01 \%. Unfortunately, this
accuracy cannot be provided in monitoring observations during
searches for microlensing events. However, observers have already
accumulated considerable experience in triggering large telescopes
to observe detected events involving microlensing by gravitational-
lens caustics, so that prospects for observing these polarization
effects are fairly encouraging.

\section{Microlensing of Stars by NUT Gravimagnetic Lenses}

Newman, Unti, and Tamburino (NUT, 1963) considered a space-time
metric that is a generalization of the Schwarzschild metric. Along
with the mass of ordinary matter, $M$, it is also determined by the
so-called NUT factor, $l$, which describes the contribution of
magnetic monopole to the curvature of space-time. The parameter $l$
is often simply called the magnetic mass. The NUT metric becomes
the Schwarzschild metric for $l \rightarrow 0$. On the other hand, gravitation
still exists as $M \rightarrow 0$ (the "pure NUT" case).

      The ability of the NUT objects to be gravitational lenses was
studied by \cite{nou97}. Generally, the
properties of such lenses are close to those of Schwarzschild
lenses. Under certain conditions, they are also capable of forming
multiple images of lensed objects and amplifying the light flux
detected by the observer. However, there are two important
differences: the presence of a round zone where the lensed source
is invisible, with its center towards the lens and its size
increasing with the NUT factor, $l$; and a shear of the source
images, manifests as a rotation in the plane of the sky around the
optical axis of the lens.

      It is possible that the angular separation between the source
images formed by a NUT lens will be smaller than a telescope's
resolution. In this case, we are dealing with gravitational
microlensing, and the observer will detect only variations of the
light flux during the lens's motion relative to the source. This
situation was analyzed for NUT lensing of a point source in papers
of \cite{rah03} and \cite{rah04}, where the results were also
compared to lensing by a Schwarzschild lens.

      The amplification coefficient for the flux from a point source
lensed by a Schwarzschild lens is given by \citep{koc06, zak97}
$$
A(u)=\frac{u^2+2}{u\sqrt{u^2+4}}. \eqno(17)
$$
Here as earlier, $u = p/p_0$ is the angular distance between the lens
and the source measured in Einstein cone angular radii. When
describing microlensing by a NUT object, a new parameter is
introduced:
$$
N = \frac{1}{p_0} \sqrt{\frac{2lD_{ds}}{D_d D_s}}.
$$
When $u^2 < 2(\sqrt{4N^4+1}-1)$ , the amplification coefficient of a NUT lens is
imaginary, corresponding to an absence of source images. Outside
this region, the amplification coefficient for the flux from an
infinitely small source is given by \cite{rah03}
$$
\begin{array}{cc}
A(u)= \frac{1}{1-a_-} \left (1-\frac{8N^4}{a_-^{-1}-1}\right )^{-1/2}-\\
\frac{1}{1-a_+} \left (1-\frac{8N^4}{a_+^{-1}-1}\right )^{-1/2} ,
\end{array}
\eqno(18)
$$
where
$$
a_\pm=\left[\frac{2+u^2\pm\sqrt{u^4+4u^2-16N^4}}{2(4N^4+1)}\right]^2
$$
and the sign in the subscript of a corresponds to the sign in front
of the square root. An approximate expression for the amplification
coefficient valid for small $N$ and $A(u)$ is
$$
A(u)=\frac{u^2+2}{u\sqrt{u^2+4}}+\frac{8N^4(u^2+2)}{u^3(u^2+4)^{3/2}}.
\eqno(19)
$$
Comparing (19) and (17), we note that amplification by a NUT lens
is stronger than amplification by a Schwarzschild lens when
$u^2 > 2(\sqrt{4N^4+1}-1)$ . With decreasing magnetic mass $(N \rightarrow 0)$,
the second
term in (19) vanishes, and this equation becomes equal to (17). The
presence of the source invisibility zone for $N$ not equal to zero
limits the minimum value of $u$, and decreases the maximum possible
amplification coefficient for a NUT lens, compared to a
Schwarzschild lens.

      In a microlensing event for a real star with finite angular
size, the flux from each infinitesimal area of its disk will be
multiplied by the amplification coefficient (17) or (19), depending
on the lens type. Let us assume for simplicity that the brightness
distribution across the stellar disk, $b_\lambda$, which depends on the
wavelength $\lambda$, can be represented with a linear law (9). Let $r$ be
the stellar angular radius and $p$ the angular distance from the
stellar disk center measured in Einstein angular radii $p_0$. The
brightness distribution (9) can then be written
$$
b_\lambda(p)=b_\lambda^0(1-x_\lambda+x_\lambda\sqrt{r^2-p^2}/p),\eqno(20)
$$
where $b_\lambda^0$ the brightness at the center of the stellar disk.

      Searches for microlensing events are based on stellar
photometry in a system close to the standard broadband $UBVRI$
system. We can write expressions similar to (9) and (20) for the
brightness distributions across a stellar disk in these filters,
introducing the brightness at the disk center and the limb-darkening
coefficient for a given filter, for example, $b_V^0$  and $x_V$
for the $V$ filter. We obtain the flux in this filter in the absence
of the lens, $H_V^0$ , by integrating the brightness distribution over
the entire visible stellar disk:
$$
H_V=\pi r^2 b_V^0 (1-x_V/3). \eqno(21)
$$
The limb-darkening coefficients in the filters of the standard
broadband system were computed by  \cite{rub90} for stars
with various effective temperatures using Kurucz' grid of model
atmospheres.

      In the presence of a gravitational lens at an angular distance
$u$ from the center of the stellar disk, the $V$ flux can be written
$$
\begin{array}{cc}
H_V(u)=\int\limits_{0}^{2\pi} d\varphi \int\limits_{0}^{r}
A(\sqrt{u^2+\xi^2 -2 u\xi\cos\varphi})\times \\
b_V^0(1-x_V+x_V\sqrt{r^2-\xi^2}/r)\xi\ d\xi   , \end{array}
\eqno(22)
$$
where the angle $\varphi$  is measured from the direction towards the lens
and the amplification coefficient $A$ is given by (17) for a
Schwarzschild lens or (18) for a NUT lens. Similar expressions can
also be written for the fluxes in other filters. Taking into
account the dependence of $u$ on time, $t$,
$$
u(t)=\sqrt{u_0^2+\frac{(t-t_0)^2}{t_E^2}}   ,
$$
where $u_0$ is the impact parameter - the minimum angular distance,
achieved at the moment $t_0$, and $t_E$ is the time needed to cross the
Einstein cone angular radius, we find the observed light curve, $I_V(t)$:
$$
I_V(t)=H_V(t)/H_V^0 . \eqno(23)
$$

      As where mentioned in section 3, the amplification coefficient
for any gravitational lens is independent of wavelength. However,
in the most probable case when the lens is not projected onto the
stellar disk, the rapid decrease of $A(u)$ with increasing distance,
$u$, leads to an increase in the fraction of light from the disk limb
in the observed flux. For the $UBVRI$ filters, the limb darkening
decreases with increasing maximum sensitivity wavelength. Thus, a
chromatic effect, a kind of reddening, should be observed for real
stellar microlensing events. This chromatic microlensing effect can
be revealed by measuring color indices, for example,
$$
B-V=2.5\lg(H_V/H_B)+C_{B-V}
$$
or
$$
V-R=2.5\lg(H_R/H_V)+C_{V-R} ,
$$
with the constants $C_{B-V}$ and $C_{V-R}$ determined by the zero points of
the photometric system. In the case of reddened star light, the
value of the color indices will exceed the values in the absence of
the lens, $(B-V)_0$ and $(V-R)_0$, and should be largest when the
lensed star achieves its maximum light flux.

      We computed the photometric and chromatic effects of
microlensing a star by a NUT lens assuming, as in paper of \cite{rah03} and
\cite{rah04}, that $0 \leq N \leq
0.5$. We chose a main-sequence $K0V$ star
with an effective temperature of $T_{eff} = 4900\ K$ and color indices
$(B-V)_0 = 0.890$, $(V-R)_0 = 0.740$ \citep{all73}. According to
\cite{rub90}, the limb-darkening coefficients for such a
star derived from the condition of flux conservation are $x_B =
0.859$, $x_V = 0.702$, and $x_R = 0.585$. We selected the brightnesses at
the disk center so that the computed fluxes (21) in the absence of
the lens would give the original color indices, $(B-V)_0$ and $(V-
R)_0$. We assumed the most probable case of lensing, when the NUT-
lens invisibility zone does not overlap with the stellar disk. When
$N = 0.5$, the light from the source cannot be detected when $u <
0.486$. We estimated the integrated fluxes (22) numerically with a
relative uncertainty of $10^{-4}$.

      Figure 5 displays the $V$ light curves for a star with $r = 0.01$,
barely different from a point source, with time measured in units
of the Einstein cone angular radius crossing time, $t_E$, for $u_0 = 0.6$
and $t_0 = 0$. The curves are symmetrical about the maximum flux, and
we show only their positive branches. The curve for $N = 0.5$ (solid)
is considerably different for the Schwarzschild-lens curve, plotted
as circles. In agreement with (19), the observed differences
between the lenses quickly decrease with decreasing $N$. The light
curve for $N = 0.3$, plotted as dots, already displays little
difference from the curve for a Schwarzschild lens, with the
difference becoming virtually undetectable on the scale of the
figure when $N = 0.1$. No chromatic effects are observable for any $N$
values in the studied range, $0 \leq N \leq 0.5$. Deviations of the star's
color indices from $(B-V)_0$ and $(V-R)_0$ are within the expected
calculation uncertainties.

\begin{figure} [t]
\plotone{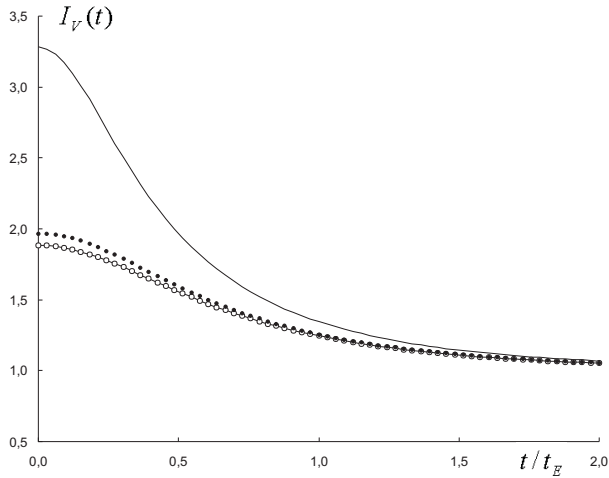}
\caption[]{
Light curves for microlensing of a star with angular
radius of $r = 0.01$ by NUT lenses with $N = 0.5$ (solid curve)
and $N = 0.3$ (dots), and by a Schwarzschild lens (circles)
}
\label{f5}
\end{figure}

      Figure 6 displays the results of our computations for a star
with $r = 0.10$, comparable to the Einstein cone angular radius, for
$u_0 = 0.6$ and $t_0 = 0$. The results for lensing of the star by a NUT
lens with $N = 0.5$ is shown by the solid curve, while the dots show
the light curve for a point source lensed by the same lens. The
curve with circles is the light curve for lensing of the same star
by a Schwarzschild lens, and virtually coincides with that for the
point source on the scale of the figure. We can see that, in
contrast to the Schwarzschild lens, the influence of the star's
finite size is fairly appreciable in the case of the NUT lens.
Thus, the difference between the $I_V(t)$ curves for the two lens
types becomes more significant as the star's angular size
increases.

\begin{figure} [t]
\plotone{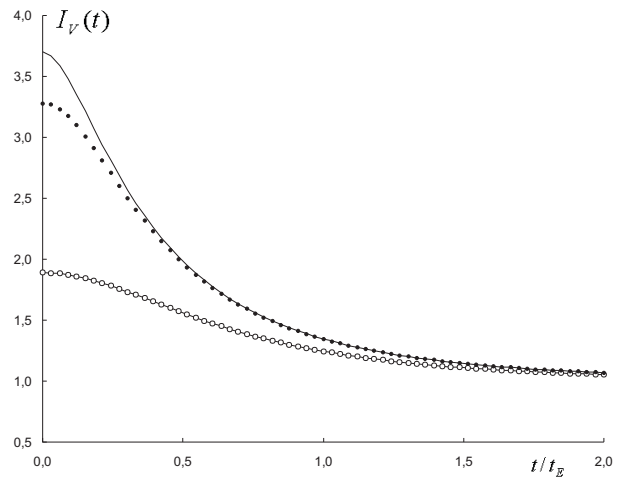}
\caption[]{
Light curves for microlensing a point source (dots)
and a star with angular radius $r = 0.10$ (solid curve) by a NUT
lens with $N = 0.5$. Also shown is the light curve for
microlensing of the same star by a Schwarzschild lens (circles)
}
\label{f6}
\end{figure}

      Figure 7 shows variations of the $B-V$ and $V-R$ color indices
for the previous lensing case ($r = 0.10$, $u_0 = 0.6$, and $t_0 = 0$). The
solid curves show the results for the NUT lens ($N = 0.5$) and the
dots the results for a Schwarzschild lens. Despite the smallness of
the chromatic effect, it is significant for the NUT lens, whereas
the star's color indices during lensing by the Schwarzschild lens
are virtually undistinguishable from the initial values, $(B-V)_0$
and $(V-R)_0$. Obviously, the chromatic effect can be made stronger
by a special choice of narrow observed spectral intervals. However,
patrol stellar microlensing observations use broadband systems.

\begin{figure} [t]
\plotone{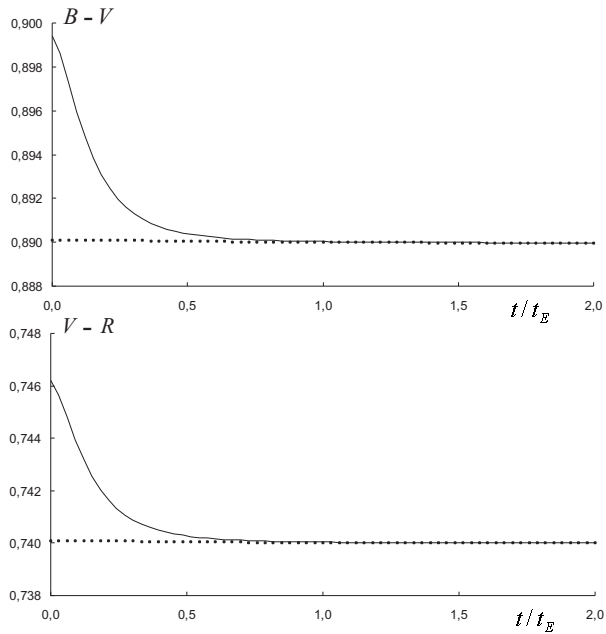}
\caption[]{
Variations of the $B-V$ and $V-R$ color indices during
microlensing of a star with angular radius $r = 0.10$ by a NUT
lens with $N = 0.5$ (solid curve) and by a Schwarzschild lens (dots)
}
\label{f7}
\end{figure}

      Another observable effect related to the predominant
amplification of light from the disk limb can be manifest during
stellar microlensing: linear polarization of the light can arise.
The larger amplification coefficient of a NUT lens for fixed $u$ in
the source visibility zone enhances this effect compared to a
Schwarzschild lens. On the other hand, it causes a shear of the
source images, rotating them in the plane of the sky; this should
give rise to a rotation of the polarization plane, depolarizing the
combined light flux. The simple approach to estimating this effect
used in section 3 is not applicable here, and this problem must be
resolved by analyzing the polarization properties of stellar images
computed using ray tracing.

\section{Conclusion}

We assume that small-scale clusters of weakly interacting
massive particles can, in the course of their evolution, acquire
rotational momentum and spheroidal shape. Even a small degree of
oblateness similar to that of the critical surface for a Roche
model can result in the appearance of caustics in the plane of a
lensed source. The multiple images that form cannot be resolved
using current telescopes, and the cluster is manifest as a non-
compact, spheroidal gravitational microlens. The crossing of
caustics in the case of relative motion of the observer, particle
cluster, and lensed star can produce a large variety of flux
curves, including some observed in actual microlensing events that
have been interpreted as manifestations of binary gravitational
lenses. We must therefore bear in mind the possibility that at
least some such events could in reality be associated with WIMP
clusters.

      The unambiguous identification of such an exotic object as a
wormhole from observations of gravitational microlensing of stars
is a rather complicated problem. If a source with a small angular
size does not cross the circular caustic, it is impossible to
discriminate between gravitational lenses with positive and
negative mass. Even for stars with relatively large angular
diameters, deviations from the light curve for a Schwarzschild lens
are of the order of the observational errors. Distinctive
photometric features of microlensing by a wormhole are a sharp
decrease of the flux considerably below the initial level during
the caustic crossing and a symmetric light curve. However,
unavoidable gaps in the observations may prevent the detection of
the second caustic crossing, and the shape of the light curve in
the vicinity of the maximum is approximately the same as for
microlensing by a caustic of an ordinary binary gravitational lens.
When the angular size of the star is comparable to $0.10$ of
the Einstein cone radius for sufficiently large impact parameters, the
observed light curves should resemble those for microlensing of a
compact object surround¥d by an extended gaseous envelope. This
case can be easily distinguished from a negative-mass lens through
an analysis of the chromatic and polarization effects arising
during the lensing.

      Our results demonstrate also that, during gravitational
microlensing of stars, under certain conditions, the influence of
the star's finite size and the brightness distribution across the
stellar disk can facilitate the detection of lensing objects whose
local space-time is described with the NUT metric. Such effects are
virtually absent if the stellar angular radius is much smaller than the
Einstein cone angular radius.
If the size of the star becomes order $0.10$ of the Einstein radius,
lensing by a NUT lens exhibits considerable differences from
lensing by a Schwarzschild lens, in terms of both flux
amplification and color-index variations.

\acknowledgments
This work was partially supported by the Russian Foundation for
Basic Research, the Ministry of Science and Education of the
Russian Federation, and the project "Universities of Russia".

\nocite{*}

\end{document}